# High-Throughput Exploration of Refractory High-Entropy Alloys for Strength and Plasticity

Stephen A. Giles[1], Hugh Shortt[2], Peter K. Liaw[2], Debasis Sengupta[1]

[1] CFD Research Corporation, 6820 Moquin Drive NW, Huntsville, AL 35806

[2] Department of Materials Science and Engineering, The University of Tennessee, Knoxville, TN 37996.

**One Sentence Summary:** Researchers leveraged deep learning to discover new high-entropy alloys with exceptional strength and plasticity.

**Abstract**

Refractory high-entropy alloys (RHEAs) are compositionally complex materials which have been demonstrated to have the potential for exceptional strength at high operating temperatures. However, their composition space is vast, and other property requirements, such as acceptable plasticity at room-temperature, must be met. Here, we leverage recently published, state-of-the-art deep learning models to predict compressive yield strength at 1,000 °C and room-temperature plasticity of >100,000 RHEAs. Multiple candidate materials were identified which exhibited exceptional balance between strength and plasticity. Upon experimental synthesis, multiple candidates were proven to outperform any previously reported RHEAs for simultaneous strength and plasticity. Our work demonstrates the power of data-driven approaches for rapid materials design, and enables continued multi-property optimization and materials discovery.

**Introduction**

Design of new, revolutionary materials frequently depend on their meeting multiple requirements and optimizing engineering trade-offs. Principles of materials design have traditionally been rooted in knowledge of underlying physics and trial-and-error experimentation. For many materials design problems, however, material properties are influenced by a myriad of factors which render a strictly physics-driven approach infeasible. As new, more complex materials are engineered to satisfy multiple design requirements, the underlying physics become more complex and, often, elusive. High-entropy alloys (HEAs) (*1*, *2*), sometimes referred to as multi-principal element alloys (MPEAs) or complex concentrated alloys (CCAs), are one such class of recent materials which have been proven capable of exhibiting excellent mechanical properties. Especially for high-temperature conditions (e.g., beyond 1,600 °C), such as in next-generation gas engine turbines (*3*), refractory HEAs (RHEAs) have emerged as promising candidate materials (*4*, *5*). The multi-principal element nature of RHEAs, and other similar materials, widens the available composition space, but severely complicates rational design of new materials (*6*).

Recently, materials design efforts have become increasingly focused on simultaneous property improvement. Arguably, the most important design problem within the RHEA space is engineering for both excellent high-temperature strength and high ductility at room-temperature, with numerous experimental and computational studies being devoted to addressing this challenge (*7*–*13*). Rao *et al.* (*14*) have used the edge dislocation model developed by Maresca *et al.* (*15*), together with the valence electron count (VEC) as a surrogate for ductility, to identify RHEAs with a good balance of strength and ductility. Tandoc *et al.* (*16*) have performed high-throughput screening of lattice distortion, strength, and intrinsic ductility of RHEAs, wherein they have leveraged the VEC as a surrogate for the *actual* ductility, in a similar fashion to Rao *et al.* (*14*) In a recent study, we have developed an optimization framework for improving the temperature-dependent yield strength of RHEAs (*17*). We recently expanded the scope to include prediction of room-temperature plasticity, uncertainty quantification, and model interpretability (*18*), as well as investigating graph-based descriptions of alloys and their properties. These nascent developments have laid the groundwork for high-throughput computational studies focused on simultaneous improvement of multiple properties (*19–21*).

In this study, we leverage a group of ensemble models developed in our recent work (*18*) to perform high-throughput predictions yield strength and plasticity of more than 100,000 RHEAs. A graph-based hybrid model approach was developed to calculate prediction-specific weights among the model ensemble. The hybrid model was validated against an independent set of literature data and demonstrated to reduce error. The hybrid model was used to generate temperature-dependent compressive yield strength and room-temperature plasticity predictions. A selection of six RHEAs (three novel equiatomics, and three novel *non*-equiatomics) were selected for experimental synthesis and testing. Multiple RHEAs within the six candidates selected were found to exhibit exceptional balance between strength and plasticity. Among these, CrMoNbTaV, a never-before-reported equiatomic RHEA, was found to have a yield strength of 1330 MPa at 900 °C and a room-temperature plasticity of 31.7%. Likewise, a non-equiatomic $Hf_{25.6}Nb_{25.1}Ta_{22.6}V_{22.3}W_{4.5}$ was found to possess a yield strength of 890 MPa at 900 °C and a room-temperature plasticity of greater than 40%, having never underwent fracture. To our knowledge, these combination s of high-temperature strength and plasticity far exceeds any HEAs which have been reported to date.

**Results and discussion**

*Hybrid model for yield strength and plasticity*

The RHEA discovery framework we have leveraged in our current work is provided in Fig. 1. In our previous work, we have demonstrated the development of a random forest model of temperature-dependent, compressive yield strength, as well as neural network, random forest, and gradient boosting ensemble models for both yield strength and plasticity. In our previous work, the performance of the neural network, random forest, and gradient boosting ensembles were compared through a variety of statistical validation approaches, including testing on a statistically significant number of recently reported, unseen data. Here, we expand upon this work by proposing a graph-based hybrid model approach to assign weights to the neural network, random forest, and gradient boosting ensembles. The hybrid model returns a single mean value, with corresponding uncertainty, for each new prediction. In addition to the novel graph-based hybrid model approach, the major novelty of our approach is its high-throughput exploration of the RHEA composition space. More than 100,000 RHEAs are selected through constrained Latin Hypercube sampling of the 10-element composition space. The high-temperature (1,000 °C) compressive yield strength and the room-temperature plasticity for each RHEA are predicted to assess the strength-plasticity trade-off for each of the RHEA compositions. Selection of candidate RHEAs is performed through figure-of-merit calculated based on Pareto optimization. Finally, a selection of six RHEAs are chosen for experimental validation.

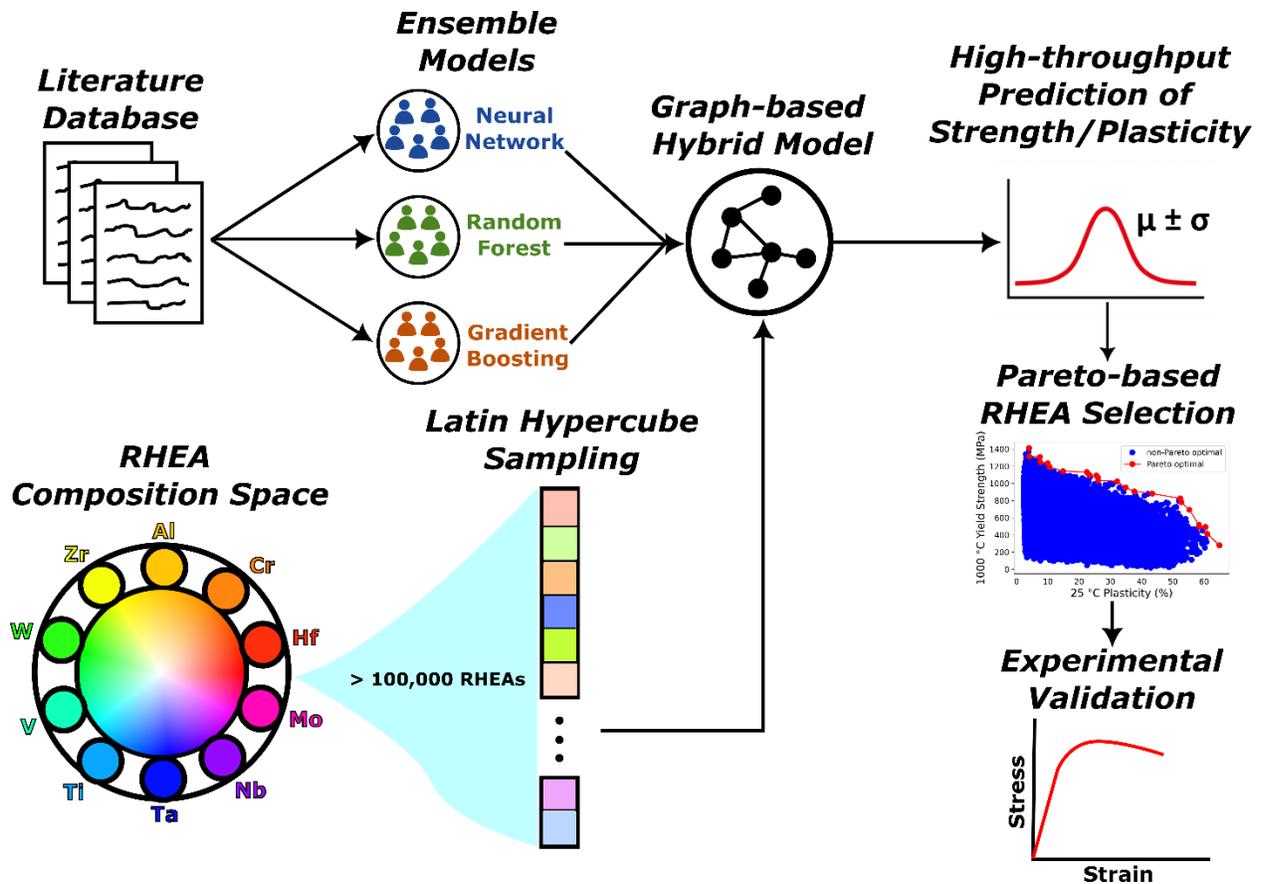

**Fig. 1. Approach overview.** In previous work (*17*), we have developed a machine learning framework based on yield strength and plasticity datasets collected by Couzinie *et al.* (*22*), Borg *et al.* (*23*), as well as our own literature data collection and generation through experimental synthesis. First, a graph-based hybrid model is employed to decide how to weight the individual ensemble predictions of the neural network, random forest, and gradient boosting models which were developed in our previous work (*18*). Next, compositions belonging the 10-element RHEA composition space are generated through Latin hypercube sampling. Predictions of the 1,000 °C yield strength and room-temperature plasticity are made for each of these compositions, eventually leading to selection of candidate RHEAs and experimental validation.

Within our framework, we have developed ensembles of neural network, random forest, and gradient boosting models (i.e., an ensemble of 100 individual models for each model architecture) (*18*). While each of these models demonstrated skill at predicting both temperature-dependent yield strength and room-temperature plasticity, it is well-known that different model architectures have various strengths and weaknesses, including a strong learner's (e.g., neural network) potential to overfit, and the lack of extrapolation capabilities afforded by tree-based models. To achieve the best overall model performance, we developed a novel graph-based hybrid model approach to select how to weight each of the model ensembles on the basis of similarity to known alloys with known model prediction errors. Dimensionality reduction of the input features (Fig. 2, A and B) was performed through isomap embedding (*24*) for the yield strength and plasticity, respectively, to quantify the similarity, in terms of Euclidean distance, of alloys in the validation set to known alloys. The results provided here for the isomap embedding represent the best clustering achieved with respect to the performance of the individual model ensembles. The effect of hyperparameters on the isomap is provided in fig. S1. To calculate the hybrid model prediction, the distance of the new alloy is calculated with respect to its nearest *k* neighbors. The distance to each of its neighbors, combined with the best-performing model for each neighbor, are then used to determine how to weight each of the three ensemble models for a new alloy (Fig. 2C). As shown in Fig. 2D, our hybrid model succeeds in reducing the prediction error for unseen data. A parity plot of the hybrid predictions in comparison to each of the three model ensembles reveals that the hybrid model significantly reduces the number of outlier predictions (fig. S2). Interestingly, as $n_{\text{neighbors}} \rightarrow \infty$, the hybrid model prediction approaches that of a simple average amongst the neural network, random forest, and gradient boosting ensemble predictions, as one might expect (fig. S3). A detailed example of calculating the hybrid model prediction for a given alloy is provided in the Supplementary Information.

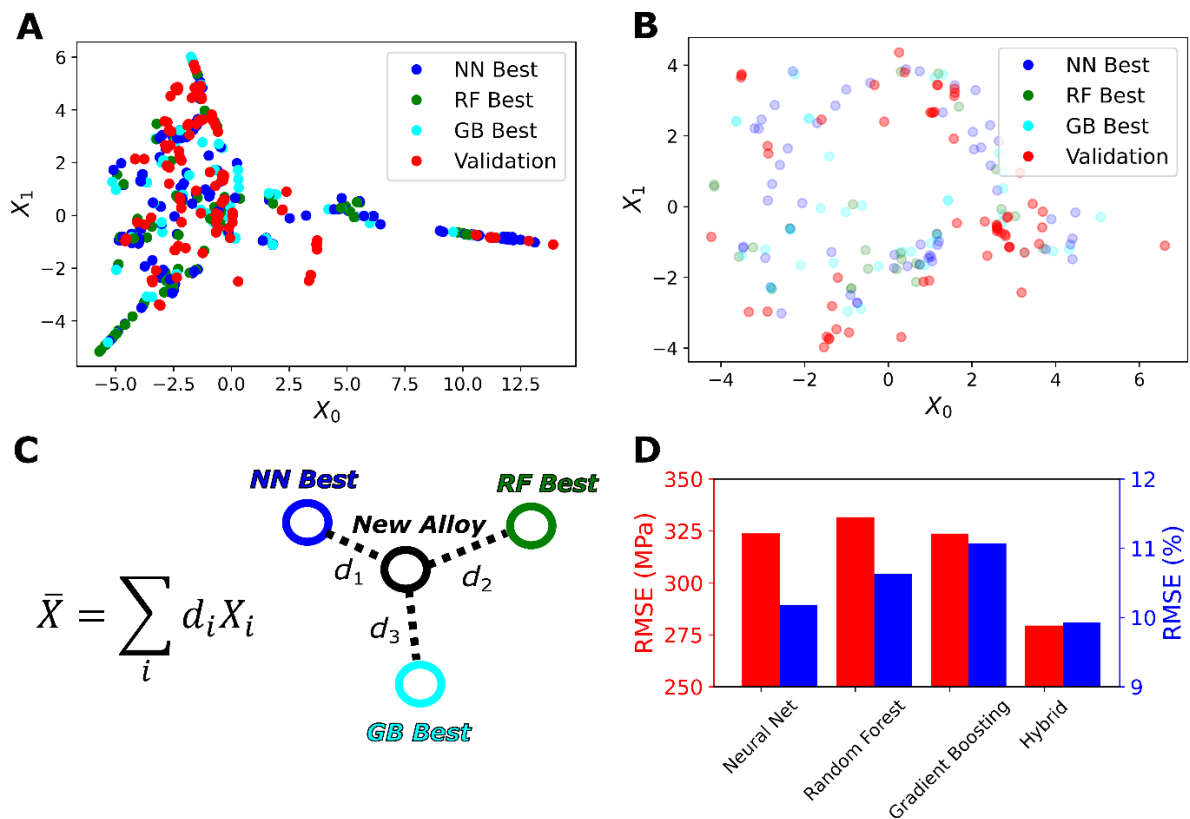

**Fig. 2. Graph-based hybrid model approach for aggregating ensemble model predictions.** (**A** and **B**) Isomap projections onto two-dimensional space of the input variables for the compressive yield strength and room-temperature plasticity datasets. (**C**) Schematic of the approach for weighting ensemble model predictions according to where a new alloy is located relative to known alloys where a (**D**) Summary of validation error reduction achieved for both yield strength and plasticity.

*High-throughput prediction of mechanical properties*

Using the hybrid model framework, the RHEA composition space can be rapidly explored for the screening of high-temperature yield strength and room-temperature plasticity. We confine the exploration to the ten elements which most frequently appear in RHEAs, namely Al, Cr, Hf, Mo, Nb, Ta, Ti, V, W, and Zr. Likewise, to mimic typical RHEA compositions, we confine the exploration to include only compositions containing 4 – 8 principal elements. This results in 837 unique equiatomic alloy compositions. For each equiatomic base alloy, 120 *non*-equiatomic compositions were also randomly chosen through a constrained Latin hypercube sampling approach (*25*), resulting in a total of 101,277 RHEA compositions screened in this study. Further details are provided in the Supplementary Information. The resulting distribution of the configurational entropy ($\Delta S_{\text{mix}}/R$) is provided in Fig. 3A, and indicates the compositions correspond to high-entropy and "medium"-entropy alloys, which have garnered recent attention (*10*, *26*, *27*). The predicted distributions of the 1,000 °C compressive yield strength and the room-temperature plasticity are provided in Fig. 3, A and B, respectively. The 1,000 °C yield strength is normally distributed with mean and standard deviation of (656 ± 210) MPa. The room-temperature plasticity is heavily skewed towards values near zero, in agreement with experimental observations, with a median value of 5.9%, and with 68% of the compositions having a plasticity less than 10%. The yield strength and plasticity of all RHEA compositions (both the 837 equiatomics, as well as the >100,000 non-equiatomics) are shown in Fig. 3C. From this result, the yield strength and plasticity models are proven to capture the experimentally observed trade-off between the two properties, which should lend confidence in the physical implications of the models. Both equiatomic and non-equiatomic RHEAs are observed to follow the trade-off, with the highest fraction of RHEAs localized at a 1,000 °C yield strength of 500 – 1,000 MPa and a plasticity of less than 10% (Fig. 3D). Our model, however, also reveals striking opportunities for simultaneously enhancing both strength and plasticity, with some RHEAs predicted to possess a 1,000 °C yield strength of ~1,000 MPa and plasticity of ~30%. The ensemble nature of the models lends itself to embedded uncertainty quantification, which can be used within our framework as a secondary metric by which to select candidate RHEAs for synthesis. High-throughput computations of the room-temperature yield strength were also performed, revealing a similar trade-off with the room-temperature plasticity (fig. S4), as well as a strong correlation with the 1,000 °C yield strength (Pearson coefficient of 0.71, see fig. S5). The correlation between the room-temperature yield strength and the 1,000 °C yield strength reveals that alloys with smaller atomic size mismatch tend to retain more of their room-temperature strength at high strength, in agreement with the athermal plateau observed for screw-controlled alloys by Beyerlein *et al.* Each of the members of the hybrid model (neural network, random forest, and gradient boosting ensembles) complete predictions for all ~$10^5$ RHEAs in only a few minutes of runtime on a single CPU.

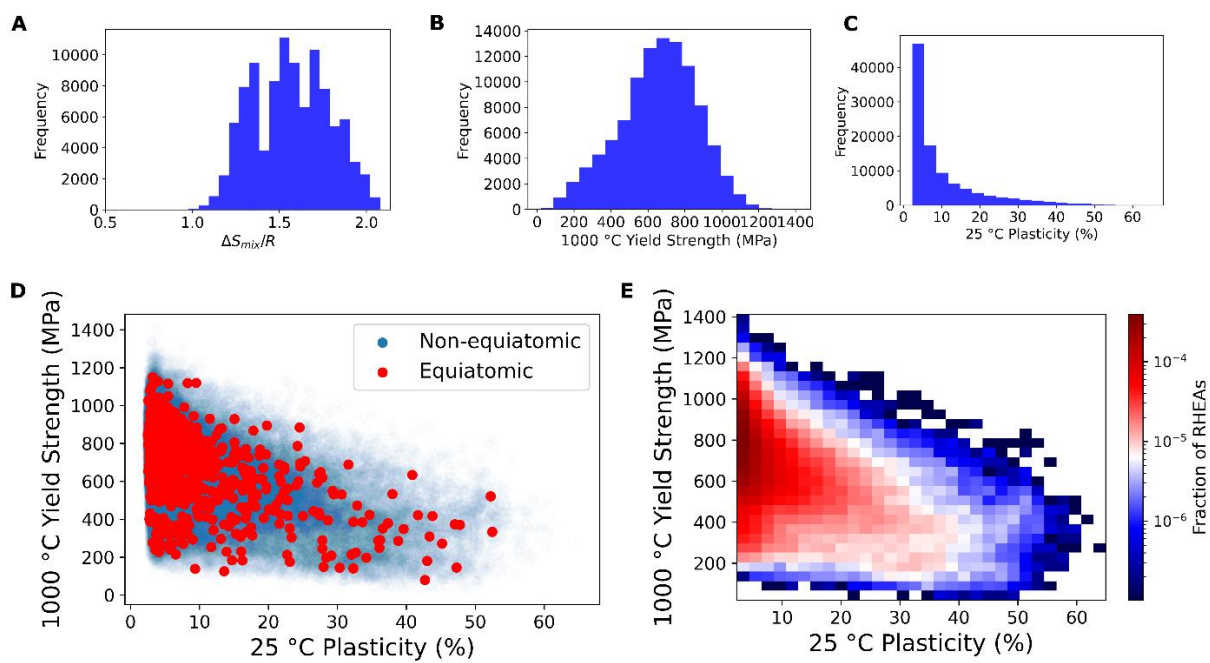

**Fig. 3. High-throughput predictions of high-temperature yield strength and room-temperature plasticity.** (**A**) Distribution of the configurational entropy for the >100,000 RHEA compositions predicted. (**B**) Distribution of the 1,000 °C compressive yield strength. The calculations are done assuming an as-cast RHEA. (**C**) Distribution of the room-temperature (25 °C) plasticity. Calculations assume an as-cast RHEA tested at a strain rate of $10^{-3}$ s$^{-1}$. (**D**) Trade-off between 1,000 °C yield strength and room-temperature (25 °C) plasticity predicted by the respective yield strength and plasticity hybrid models. (**E**) The data from panel (**D**) depicted as a two-dimensional histogram showing the location of RHEAs with respect to the high-temperature yield strength and room-temperature plasticity trade-off.

*Compositional and physics-based trends*

To rationalize our model predictions and to condense our observations into a small set of design rules, we have analyzed compositional and physical trends which foster the best balance of high-

-9-

temperature yield strength and room-temperature plasticity. Fig. 4A identifies the RHEAs according to the presence or absence of each of the ten principal elements. For a given yield strength, the data located furthest to the right (i.e., highest plasticity) define the boundary of the trade-off. Surprisingly, despite the huge variety of compositions sampled, an easily identifiable trend for each element with respect to the strength-plasticity trade-off is usually present. For example, Nb, Ta, and V are each shown to be present in the vast majority of RHEAs which lie near the trade-off boundary. In contrast, Zr and W tend to be absent from the same RHEAs. Other elements such as Cr and Mo tend to be present in RHEAs which have the highest 1,000 °C yield strengths (i.e., > 1000 MPa), but tend to be absent from RHEAs which have the highest plasticities (i.e., > 30%).

With regards to the physical descriptors which serve as model input features, we highlight three key properties which our models suggest are among the most important for optimizing the trade-off (Fig. 4, B to D). Larger values of the Poisson ratio, ν, generally correspond to greater plasticity, in accordance with known physics. Likewise, the melting temperature, $T_m$, is observed to be the most consequential descriptor for determining high-temperature strength. While both of these observations may seem obvious, we emphasize that recovery of sensible and expected physics from a data-driven model is by no means guaranteed, and that our models are capable not only of "learning" features embedded in data, but are also able to uncover *physically meaningful* features. We also note that a majority of the RHEAs that lie at or near the trade-off boundary have small to moderate values of the atomic size mismatch (Fig. 4C), and are typically comprised of four or five principal elements (fig. S6). The dislocation mechanism, as captured by a recently proposed criterion based on lattice distortion, is shown to correlate with the trade-off optimality of the RHEAs. In particular, edge-controlled dislocations are shown to generally possess better trade-off properties than screw-controlled dislocations, but with the important caveat of the RHEA being a solid solution (in this case for RHEAs, a continuous BCC phase).

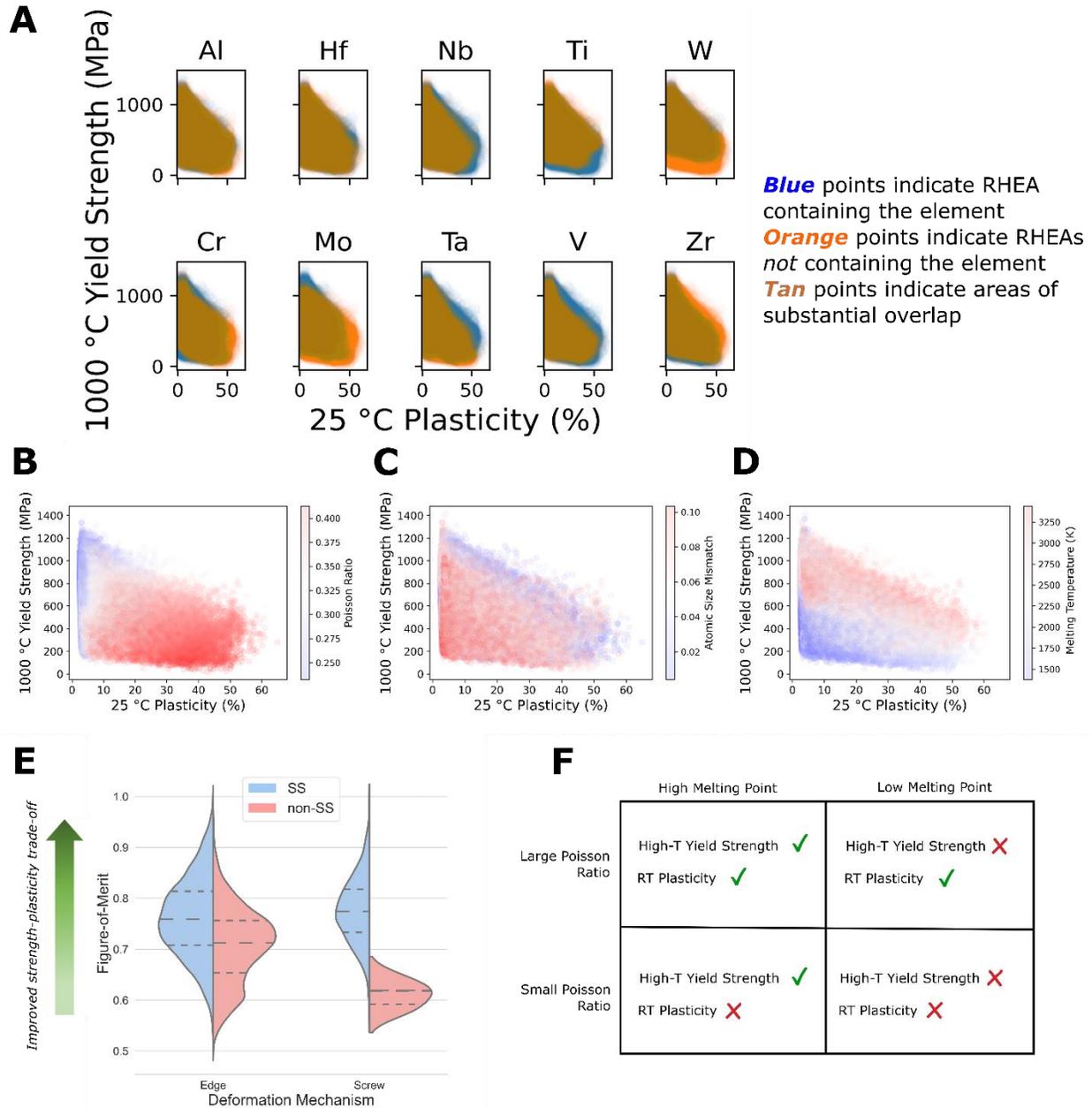

**Fig. 4. Effects of elemental and physics-based descriptors on yield strength-plasticity tradeoff.** (**A**) Trends with respect to presence of each element in the predicted RHEA compositions. (**B – D**) 1,000 °C yield strength and room-temperature plasticity predictions colored according to the RHEA's Poisson ratio, atomic size mismatch, and melting temperature. (**E**) Distribution of the figure-of-merit (closer to unity corresponds to best strength-plasticity trade-off) with respect to the deformation mechanism, separated according to their predicted phase behavior (solid-solution (SS) or non-SS). For clarity, only RHEAs with a plasticity of greater than 10% are

-11-

shown. (**F**) Simplified design rule which was found to provide reasonable guidance for developing RHEAs with high-temperature strength and room-temperature plasticity.

*Discovery, synthesis, and testing of candidate materials*

From the high-throughput dataset, the Pareto front was defining the trade-off boundary was determined (Fig. 5A). The RHEAs which comprise the Pareto front are, by definition, "Pareto-optimal", and no RHEA exists in the dataset of >100,000 compositions which has both a higher 1,000 °C yield strength *and* a higher room-temperature plasticity than a Pareto-optimal RHEA. For the remaining Pareto-inferior RHEAs, determination of the Pareto front allows for the calculation of a figure-of-merit (Fig. 5B), which is based on the distance from the Pareto-inferior RHEA to the nearest Pareto-optimal RHEA. A list of all the discovered Pareto-optimal RHEAs is provided in Table S1. The figure-of-merit was formulated such that a value of unity corresponds to a Pareto-optimal alloy, whereas a value of zero corresponds to a RHEA with no strength or plasticity (see Supplementary Information for details). The figure-of-merit quantifies how "good" or "bad" a particular RHEA is with respect to the strength-plasticity trade-off (a distribution of the figure-of-merit is provided for all RHEAs in fig. S7). The Pareto-optimal alloys were down-selected by identifying three base RHEA systems, CrMoNbTaV, AlCrNbTaV, and HfNbTaVW, which appear at multiple different non-equiatomic (NEA) compositions within the set of Pareto-optimal alloys (Fig. 5C). The NEA compositions with the best plasticities and/or lowest uncertainties in their prediction, $Cr_{12.1}Mo_{20.5}Nb_{18.6}Ta_{12.4}V_{36.5}$, $Al_{7.4}Cr_{16.6}Nb_{23.6}Ta_{23.7}V_{28.8}$, and $Hf_{25.6}Nb_{25.1}Ta_{22.6}V_{22.3}W_{4.5}$, as well as their equiatomic counterpart, were selected for experimental synthesis and characterization (details provided in the Supplementary Information) for a total of six novel RHEAs. Compressive tests are shown for 25 °C and 900 °C in Fig. 5D and E, respectively. Each of the RHEA compositions were shown to have impressive yield strengths at 900 °C, ranging from 890 MPa for NEA HfNbTaVW to 1692 MPa for AlCrNbTaV. In addition, three of the six RHEAs (CrMoNbTaV, NEA CrMoNbTaV, and NEA HfNbTaVW) were confirmed to possess a room-temperature plasticity of more than 20%, demonstrating the success of our framework in discovering new materials with exceptional trade-off properties. We provide a comparison of the RHEAs synthesized in this study with published temperature-dependent compressive yield strength and room-temperature plasticity data in Fig. 5F. In comparison to previously reported RHEAs, the CrMoNbTaV RHEA and the NEA HfNbTaVW RHEA, both reported for the first time in this study, are shown to be extremely unique, high-performing materials with respect to the typical balance of yield strength and plasticity experimentally observed. The room-temperature plasticity and 900 °C yield strength of CrMoNbTaV are measured to be 31.7% and 1330 MPa, respectively. In comparison, the room-temperature plasticity and 900 °C yield strength of NEA HfNbTaVW are measured to be >40% (i.e., the sample never underwent fracture) and 890 MPa, respectively. By contrast, as shown in Fig. 5F, previous RHEAs with comparable room-temperature plasticities were found to only have a yield strength of only ~300 MPa at the *lower* temperature of 800 °C. The RHEAs discovered through our approach maintained the high degree of plasticity, while tripling and quadrupling the yield strength at an even higher elevated temperature. Moreover, the high-throughput machine-learning framework identified these novel RHEA compositions based purely upon the currently available experimental data, without the need for further data collection and refinement by ourselves. That the outer bounds of our models' predictions with respect to the strength-plasticity can be explored, materials synthesized, and experimentally validated to possess abnormal strength-plasticity synergy should lend confidence in employing this framework for future computationally-guided materials design.

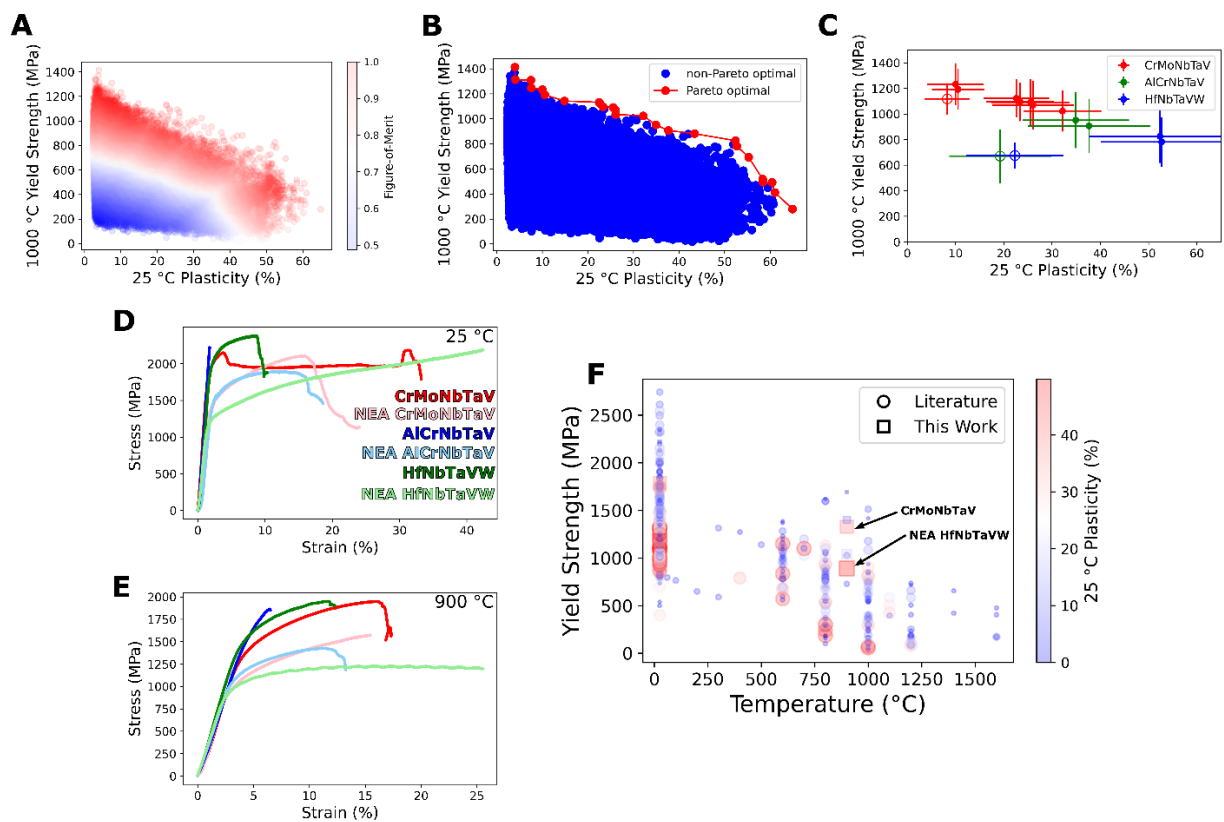

**Fig. 5. Discovery, synthesis, and testing of RHEA candidates with promising trade-off properties.** (**A**) The high-throughput RHEA predictions of 1,000 °C yield strength and room-temperature plasticity colored according to the defined figure-of-merit (where a value of unity corresponds to a Pareto-optimal alloy). (**B**) The Pareto boundary governing the yield strength-plasticity tradeoff for as-cast RHEAs determined from the high-throughput predictions. (**C**) A down-selection of non-equiatomic (NEA) Pareto-optimal alloys (shown with filled symbols) and their equiatomic, Pareto-inferior counterparts (shown with open symbols). (**D** and **E**) Stress-strain response (strain rate: $10^{-3}$ s$^{-1}$) of the six identified RHEA candidates at 25 °C and 900 °C. (**F**) Comparison of the temperature-dependent yield strength and room-temperature plasticity of the six RHEAs synthesized in this work to previous RHEA examples within the literature. Each RHEA's room-temperature plasticity is depicted by both the color and the size of the symbol, where large, red symbols correspond to high room-temperature plasticity.

**Conclusions**

Novel data-driven, graph-based hybrid models were developed to allow for rapid high-throughput materials discovery. Our data-driven modeling strategy was proven to be an effective approach for discovering RHEAs with rare, outlier properties of simultaneous strength and plasticity. Two RHEAs that were computationally identified and experimentally characterized as part of this work,

CrMoNbTaV and $Hf_{25.6}Nb_{25.1}Ta_{22.6}V_{22.3}W_{4.5}$, were each shown to have a combination of high-temperature yield strength and room-temperature plasticity heretofore unobserved. We believe that this study will serve as an important guidepost for future studies where the scope of data-driven models will further expand to include additional properties.

**References**

1. S. Ranganathan, Alloyed pleasures: Multimetallic cocktails. *Curr. Sci.* **85**, 1404–1406 (2003).

2. B. Cantor, I. T. H. Chang, P. Knight, A. J. B. Vincent, Microstructural development in equiatomic multicomponent alloys. *Mater. Sci. Eng. A*. **375–377**, 213–218 (2004).

3. J. H. Perepezko, The Hotter the Engine, the Better. *Science (80-. )*. **326**, 1068–1069 (2009).

4. D. B. Miracle, O. N. Senkov, A critical review of high entropy alloys and related concepts. *Acta Mater.* **122**, 448–511 (2017).

5. O. N. Senkov, D. B. Miracle, K. J. Chaput, J.-P. Couzinie, Development and exploration of refractory high entropy alloys—A review. *J. Mater. Res.* **33**, 3092–3128 (2018).

6. D. B. Miracle, High entropy alloys as a bold step forward in alloy development. *Nat. Commun.* **10**, 1–3 (2019).

7. Y. Zou, H. Ma, R. Spolenak, Ultrastrong ductile and stable high-entropy alloys at small scales. *Nat. Commun.* **6** (2015), doi:10.1038/ncomms8748.

8. C. Lee, G. Song, M. C. Gao, R. Feng, P. Chen, J. Brechtl, Y. Chen, K. An, W. Guo, J. D. Poplawsky, S. Li, A. T. Samaei, W. Chen, A. Hu, H. Choo, P. K. Liaw, Lattice distortion in a strong and ductile refractory high-entropy alloy. *Acta Mater.* **160**, 158–172 (2018).

9. Z. Han, J. Li, Y. Tian, A. Tian, A. Sun, R. Wei, G. Liu, Annealing-induced abnormally

ultrahigh cryogenic strength-ductility synergy in a cold rolled ferrous medium entropy alloy. *Intermetallics*. **159**, 107933 (2023).

10. N. Yurchenko, E. Panina, A. Tojibaev, S. Zherebtsov, N. Stepanov, Overcoming the strength-ductility trade-off in refractory medium-entropy alloys via controlled B2 ordering (2022), doi:10.1080/21663831.2022.2109442.

11. P. Shi, W. Ren, T. Zheng, Z. Ren, X. Hou, J. Peng, P. Hu, Y. Gao, Y. Zhong, P. K. Liaw, Enhanced strength–ductility synergy in ultrafine-grained eutectic high-entropy alloys by inheriting microstructural lamellae. *Nat. Commun.* **10** (2019), doi:10.1038/s41467-019-08460-2.

12. S. Chen, Z. H. Aitken, S. Pattamatta, Z. Wu, Z. G. Yu, D. J. Srolovitz, P. K. Liaw, Y. Zhang, Simultaneously enhancing the ultimate strength and ductility of high-entropy alloys via short-range ordering. *Nat. Commun.* **12**, 4953 (2021).

13. Y.-C. Wu, J.-L. Shao, FCC-BCC phase transformation induced simultaneous enhancement of tensile strength and ductility at high strain rate in high-entropy alloy. *Int. J. Plast.* **169**, 103730 (2023).

14. Y. Rao, C. Baruffi, A. De Luca, C. Leinenbach, W. A. Curtin, Theory-guided design of high-strength, high-melting point, ductile, low-density, single-phase BCC high entropy alloys. *Acta Mater.* **237**, 118132 (2022).

15. F. Maresca, W. A. Curtin, Mechanistic origin of high strength in refractory BCC high entropy alloys up to 1900K. *Acta Mater.* **182**, 235–249 (2020).

16. C. Tandoc, Y. Hu, L. Qi, P. K. Liaw, Mining of lattice distortion , strength , and intrinsic ductility of refractory high entropy alloys. *npj Comput. Mater.* **9**, 53 (2023).

17. S. A. Giles, D. Sengupta, S. R. Broderick, K. Rajan, Machine-learning-based intelligent

framework for discovering refractory high-entropy alloys with improved high-temperature yield strength. *npj Comput. Mater.* **8**, 235 (2022).

18. S. A. Giles, H. Shortt, P. K. Liaw, D. Sengupta, *arXiv [cond-mat]*, in press.

19. S. R. Broderick, S. A. Giles, D. Sengupta, K. Rajan, Graph-Based Data Analysis for Building Chemistry–Phase Design Rules for High Entropy Alloys. *Crystals*. **15** (2025), , doi:10.3390/cryst15010023.

20. M. T. Islam, S. A. Giles, D. Sengupta, K. Rajan, S. R. Broderick, Atomic Interaction-Based Prediction of Phase Formations in High-Entropy Alloys. *ACS Omega*. **10**, 24560–24575 (2025).

21. M. T. Islam, S. Verma, S. A. Giles, D. Sengupta, S. R. Broderick, Enhanced modeling of electronic structure – Mechanical property relationships in high-entropy alloys through data driven analysis. *Mater. Today Commun.* **47**, 113112 (2025).

22. J. P. Couzinié, O. N. Senkov, D. B. Miracle, G. Dirras, Comprehensive data compilation on the mechanical properties of refractory high-entropy alloys. *Data Br.* **21**, 1622–1641 (2018).

23. C. K. H. Borg, C. Frey, J. Moh, T. M. Pollock, S. Gorsse, D. B. Miracle, O. N. Senkov, B. Meredig, J. E. Saal, Expanded dataset of mechanical properties and observed phases of multi-principal element alloys. *Sci. Data*. **7**, 1–6 (2020).

24. J. B. Tenenbaum, V. de Silva, J. C. Langford, A Global Geometric Framework for Nonlinear Dimensionality Reduction. *Science (80-. )*. **290**, 2319–2323 (2000).

25. M. D. McKay, R. J. Beckman, W. J. Conover, A comparison of three methods for selecting values of input variables in the analysis of output from a computer code. *Technometrics*. **21**, 239–245 (1979).

26. H. Yao, J.-W. Qiao, M. C. Gao, J. A. Hawk, S.-G. Ma, H. Zhou, MoNbTaV Medium-Entropy Alloy. *Entropy*. **18** (2016), doi:10.3390/e18050189.

27. Y. C. Huang, C. H. Su, S. K. Wu, C. Lin, A study on the hall-petch relationship and grain growth kinetics in FCC-structured high/medium entropy alloys. *Entropy*. **21** (2019), doi:10.3390/e21030297.

**Acknowledgments**

We thank Prof. Krishna Rajan and Prof. Scott Broderick from the University at Buffalo for helpful conversations. *Funding:* This work was funded by the Office of Naval Research of the United States under the Small Business Technology Transfer program (Contract # N68335-20-C-0402). *Author contributions:* S.A.G., D.S., and P.K.L. conceived the study. S.A.G. and D.S. developed the software and analyzed the results. H.S. performed the experiments. S.A.G. produced the final figures. All authors discussed and commented on the manuscript.

**Competing interests**

The authors declare no competing interests.

**Data and materials availability**

The training datasets for yield strength and plasticity were curated from a combination of the Citrine MPEA dataset (*23*), the Couzinie RHEA dataset (*22*), and recent literature collected by us. The data used to develop the temperature-dependent compressive yield strength model and the room-temperature plasticity model can be found in the supplementary materials. The readers are requested to contact the authors for the code used to train the models and generate the high-throughput predictions.

**Materials and Methods**

*Calculation of hybrid model predictions and uncertainties*

The prediction value provided by the hybrid model is simply the arithmetic mean of the model ensembles, weighted according to their distance from the prediction alloy:

$$\mu_{\text{hybrid}} = \sum_i w_i \mu_i$$

The weights of individual model ensembles, $w_i$, (neural network, random forest, and gradient boosting) are normalized to sum to unity. The weights are calculated according to their distance from known alloys which are best predicted by either of the three model ensembles.

The uncertainty of the model predictions is calculated from taking the sum of the variances of the individual model ensembles, according to,

$$\sigma_{\text{hybrid}} = \sqrt{\sum_i w_i \sigma_i^2}$$

where $w_i$ is the weight (between 0 and 1) of the $i^{\text{th}}$ ensemble model (in this case, either neural network, random forest, or gradient boosting), and $\sigma_i$ is the uncertainty (i.e., the standard deviation of the bootstrap ensemble) of the $i^{\text{th}}$ ensemble model.

*Constrained Latin hypercube sampling*

We confine the exploration to the ten elements which most frequently appear in RHEAs, namely Al, Cr, Hf, Mo, Nb, Ta, Ti, V, W, and Zr. Likewise, to mimic typical RHEA compositions, we confine the exploration to include only compositions containing 4 – 8 principal elements. For each equiatomic base alloy, 120 *non*-equiatomic compositions were also randomly chosen through a constrained Latin hypercube sampling approach (*19*), resulting in a total of 101,277 RHEA compositions screened in this study. The constraints applied to the Latin hypercube sampling were defined in terms of the mole fractions of the constituent elements. For elements present in a given base alloy, the element fractions were constrained to between 20% and 200% of the corresponding equiatomic fraction. Taking the CrMoNbTaV quinary base alloy as an example, the equiatomic mole fraction for each element corresponds to 0.2. Therefore, sampling of the elemental space was constrained between element fractions of 0.04 – 0.4 for each element. This constraint was applied in order to ensure that each of the compositions which were randomly sampled were generally relevant to typical compositions in RHEAs.

*Determination of Pareto boundary*

On the basis of the model predictions of both the yield strength and the plasticity, determination of the Pareto boundary is straightforward and simply requires sorting the composition rows by either of the yield strength or plasticity prediction values. The compositions with the maximum value of yield strength and the maximum value of plasticity are both Pareto optimal compositions by definition. The plasticity value of the composition with maximum yield strength is stored in memory. Then, going in the direction of descending yield strength, the first composition encountered which has a higher plasticity value than the maximum yield strength alloy is a Pareto optimal composition (i.e., no RHEA in the high-throughput dataset has both a higher yield strength *and* a higher plasticity). The highest observed plasticity value is updated, and the process continued going in the direction of descending yield strength until the composition corresponding to the maximum plasticity in the dataset is reached. The result is a table of Pareto optimal compositions, which is provided below in Table S1.

**Table S1.** Summary of discovered Pareto-optimal RHEAs, sorted according to their mean yield strength at 1000 °C, with predicted strength (MPa) and plasticity (%), and their respective uncertainties, provided. The three NEA compositions which were selected for experimental synthesis as part of this work are highlighted.

| Composition | Mean Yield Strength (1000C) | STDEV Yield Strength (1000C) | Mean Plasticity (25C) | STDEV Plasticity (25C) |
|---|---|---|---|---|
| Al0.26123553 Mo0.27351493 Ta0.08006687 V0.11880159 W0.04395749 Zr0.22242357 | 1412.34 | 393.97 | 3.96 | 3.07 |
| Al0.25175262 Mo0.26562391 Ta0.09951421 W0.1072434 Zr0.27586586 | 1313.41 | 378.58 | 4.05 | 3.32 |
| Al0.0540978 Cr0.19057906 Mo0.19324724 Nb0.19196662 Ta0.06330919 Ti0.04455427 V0.22903734 W0.03320847 | 1308.69 | 221.24 | 7.48 | 5.19 |
| Al0.06309872 Cr0.23360453 Hf0.11812554 Mo0.04840194 Nb0.16143045 Ta0.2509432 V0.12439562 | 1247.12 | 249.37 | 7.55 | 6.02 |
| Cr0.25034551 Mo0.19868129 Nb0.16507686 Ta0.06272739 V0.32316895 | 1233.32 | 165.11 | 9.99 | 5.87 |
| Cr0.17743341 Mo0.3026388 Nb0.10860871 Ta0.06463071 V0.34668838 | 1192.47 | 159.82 | 10.52 | 5.43 |
| Al0.06418469 Cr0.24739844 Mo0.03789042 Nb0.05022917 Ta0.21690778 Ti0.1228644 V0.2605251 | 1141.92 | 215.41 | 14.83 | 8.70 |
| Al0.07073248 Cr0.17022093 Mo0.06942724 Nb0.10273682 Ta0.16715108 V0.41973146 | 1132.09 | 201.33 | 22.46 | 10.34 |
| Cr0.12053511 Mo0.20468439 Nb0.18575678 Ta0.12378189 V0.36524184 | 1123.02 | 148.07 | 22.60 | 6.79 |
| Cr0.07577973 Mo0.31225239 Nb0.15813582 Ta0.0659815 V0.38785056 | 1095.58 | 151.45 | 23.36 | 7.03 |
| Cr0.14221819 Mo0.15303159 Nb0.1185994 Ta0.1252807 V0.46087012 | 1091.84 | 183.27 | 25.53 | 8.14 |
| Cr0.14297539 Mo0.20784049 Nb0.08087643 Ta0.05954872 V0.50875897 | 1070.06 | 191.81 | 26.04 | 8.45 |
| Al0.05640107 Cr0.14410279 Nb0.0525061 Ta0.11164468 V0.44410729 W0.19123807 | 1035.30 | 215.46 | 26.08 | 15.90 |

| Composition | | | | |
|---|---|---|---|---|
| Cr0.06058604 Mo0.1940529 Nb0.23626321 Ta0.05850097 V0.45059689 | 1023.22 | 160.41 | 32.16 | 8.05 |
| Al0.07411458 Cr0.18513536 Nb0.27387299 Ta0.12216264 V0.34471443 | 951.84 | 218.49 | 34.90 | 11.06 |
| Al0.07350463 Cr0.16552911 Nb0.2360909 Ta0.23701664 V0.28785872 | 906.87 | 212.48 | 37.66 | 12.67 |
| Al0.06341992 Cr0.1474794 Ta0.23416467 V0.55493601 | 881.27 | 260.62 | 43.41 | 17.37 |
| Hf0.33230032 Nb0.14561271 Ta0.26142246 V0.21557654 W0.04508798 | 792.91 | 200.75 | 52.38 | 14.70 |
| Hf0.25554109 Nb0.2509764 Ta0.22610827 V0.22254934 W0.0448249 | 744.68 | 182.55 | 52.71 | 12.60 |
| Cr0.09794148 Nb0.24401356 Ta0.114633 V0.54341196 | 693.00 | 232.28 | 55.28 | 14.22 |
| Al0.045442 Hf0.22118053 Nb0.26825555 Ta0.18011715 V0.28500476 | 518.98 | 160.22 | 58.35 | 15.91 |
| Nb0.21524335 Ta0.17893418 Ti0.09391023 V0.51191223 | 494.77 | 163.78 | 58.44 | 13.85 |
| Al0.083103 Nb0.22942937 Ta0.18261594 V0.50485169 | 492.77 | 195.41 | 60.45 | 16.20 |
| Al0.0972784 Nb0.32722413 Ta0.08607877 V0.4894187 | 412.75 | 197.40 | 61.04 | 15.44 |
| Nb0.43234979 Ti0.42223356 V0.09582736 W0.04958929 | 279.68 | 145.68 | 64.92 | 11.95 |

*Definition and calculation of figure-of-merit*

On the basis of the set of the Pareto optimal points determined, a figure-of-merit is calculated which ranks all RHEAs in terms of their maximizing of the tradeoff. The figure-of-merit is calculated by normalizing the yield strength and plasticity axes to be between 0 and 1. Then, for every RHEA in the dataset, its distance from the nearest Pareto optimal alloy is calculated. Since all plasticity and yield strength values are between 0 and 1, the maximum possible distance is $\sqrt{2}$. Consequently, the figure-of-merit for a given RHEA is defined as,

$$FoM = 1 - \frac{\sqrt{\Delta YS_{norm}^2 + \Delta Pl_{norm}^2}}{\sqrt{2}}$$

, where $\Delta YS_{norm}$ and $\Delta Pl_{norm}$ are the different between the given RHEA's and the nearest Pareto optimal RHEA's normalized yield strength and plasticity values, respectively. Thus, the figure-of-merit will be equal to unity for a Pareto optimal RHEA, by definition.

*Experimental sample preparation and mechanical property testing.*

Ingots of CrMoNbTaV, $Cr_{12.1}Mo_{20.5}Nb_{18.6}Ta_{12.4}V_{36.5}$, AlCrNbTaV, $Al_{7.4}Cr_{16.6}Nb_{23.6}Ta_{23.7}V_{28.8}$, HfNbTaVW, and $Hf_{25.6}Nb_{25.1}Ta_{22.6}V_{22.3}W_{4.5}$ were prepared by melting high-purity elements (all element purity > 99.5 at.%) using an argon-backfilled arc melter at room temperature. All elemental components were measured to an accuracy of ± $10^{-3}$ g. The ingots were cast and then cut into cylindrical geometries [Diameter: 4 mm, Length: 6.5 - 8 mm, Aspect ratio (diameter/length): 1.5 - 2] for both room- and high-temperature (900 °C) compression testing, employing a Materials Testing System (MTS) attached with a 647 Hydraulic Wedge Grip and an MTS 653 Furnace. The strain rate was $10^{-3}$ $s^{-1}$ for compression experiments. Lastly, prior to compression tests, all samples were polished using SiC polishing paper to 1,200 grit prior to testing.

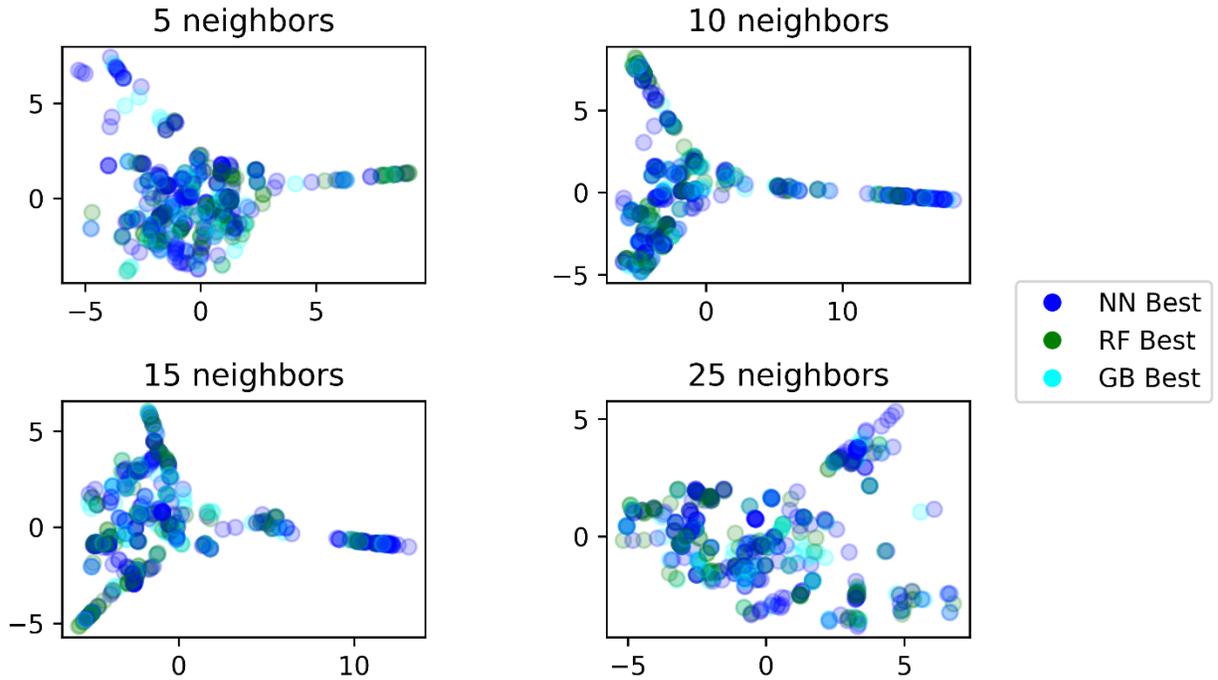

**Figure S6.** Effect of number of neighbors used to compute the isomap on the relative clustering observed.

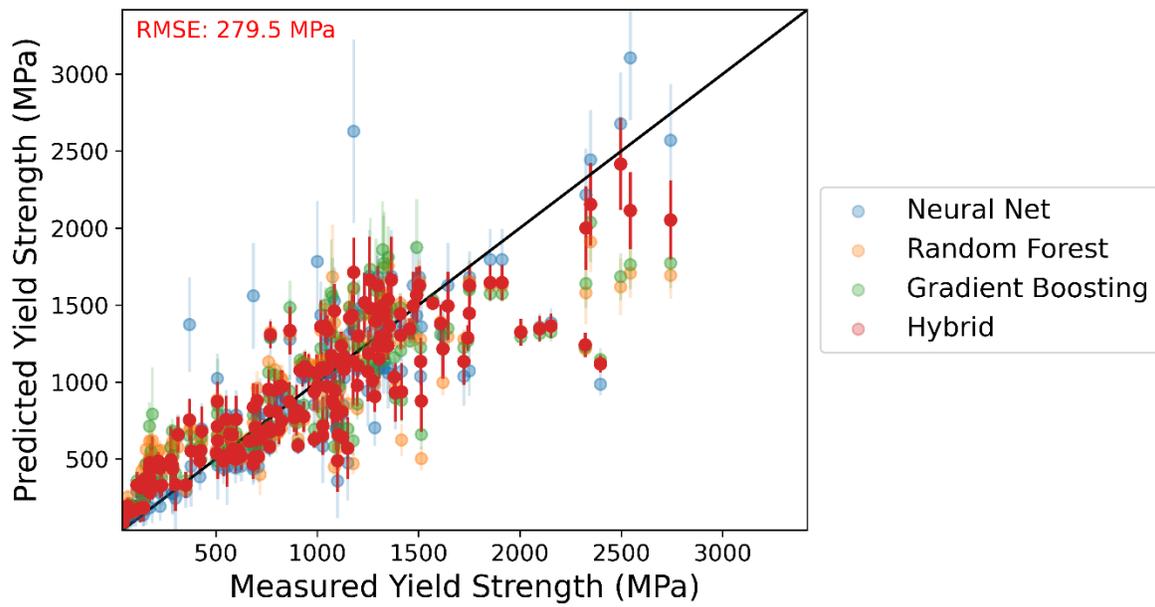

**Figure S7.** Comparison of the hybrid model predictions of the unseen validation test set to the other three ensemble model types for yield strength. The hybrid model's RMSE is significantly lower than either of the three model ensembles.

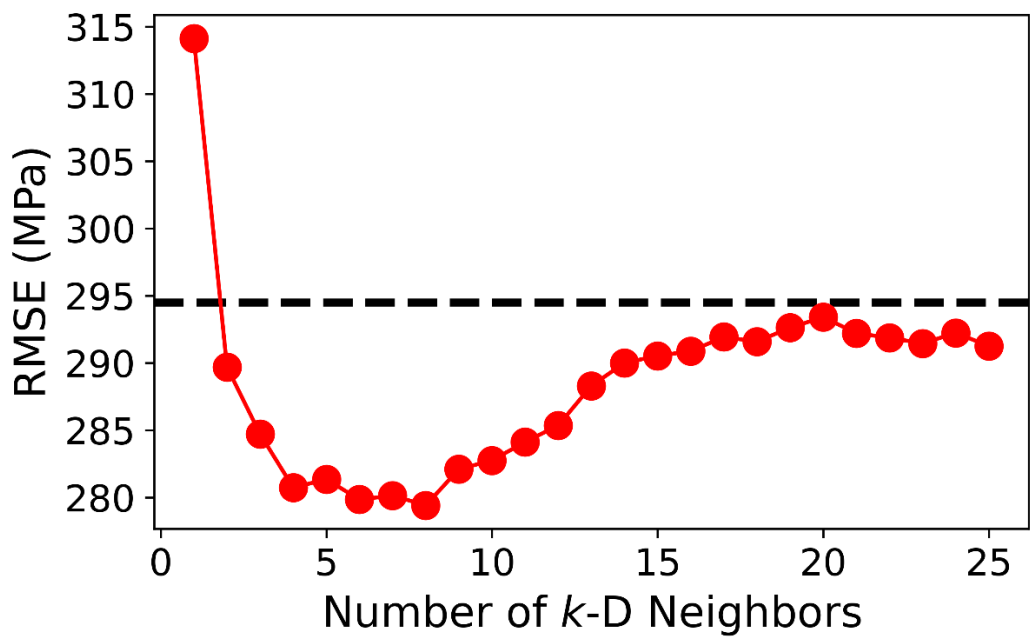

**Figure S8.** Trend of the hybrid model RMSE with respect to the number of nearest neighbors included in the model weighting scheme. The horizontal dashed line indicates the RMSE obtained from taking a simple mathematical average of the three ensemble model predictions.

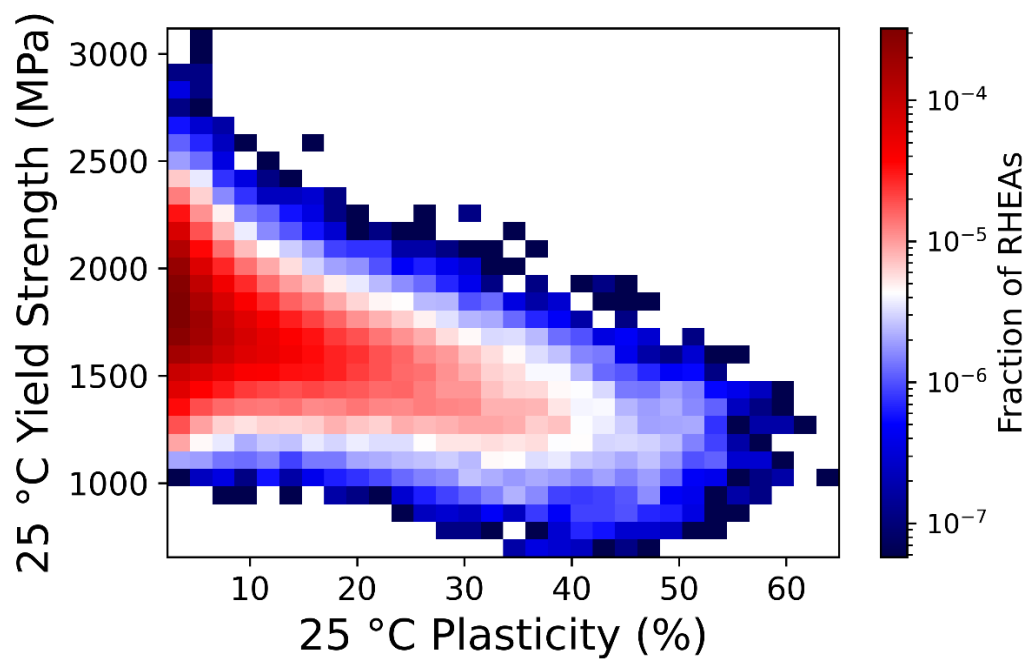

**Figure S9.** Two-dimensional histogram showing the location of RHEAs with respect to the room-temperature yield strength and room-temperature plasticity trade-off.

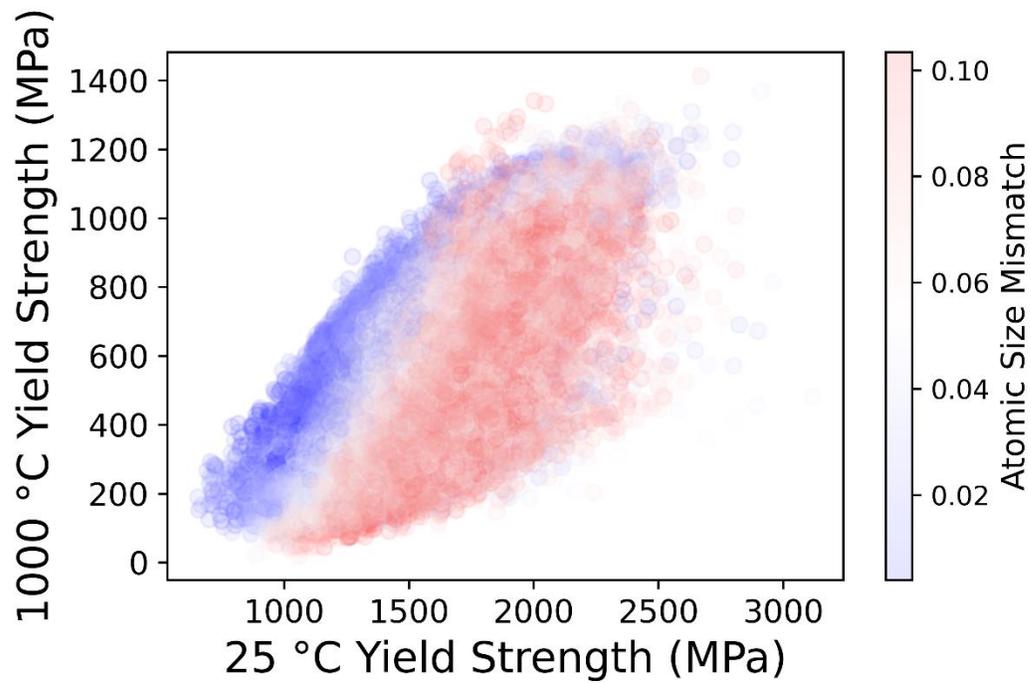

**Figure S10.** Correlation between the room-temperature yield strength and the 1,000 °C yield strength, with data colored according to the atomic size mismatch of the individual RHEAs.

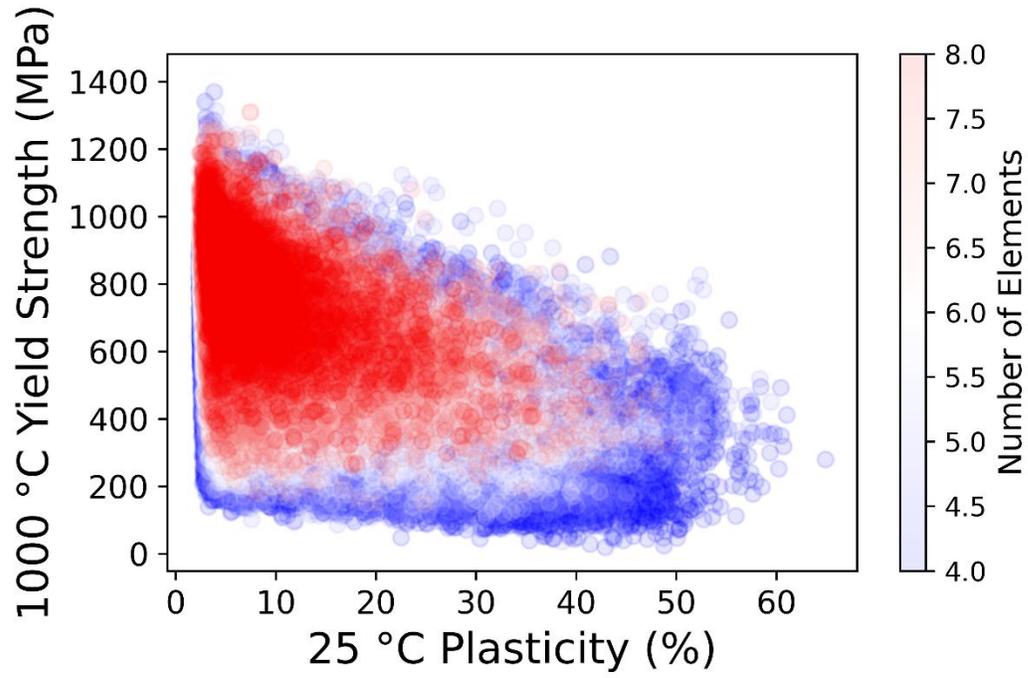

**Figure S11.** Room-temperature plasticity versus the 1,000 °C yield strength, with data colored according to the number of principal elements in the individual RHEAs.

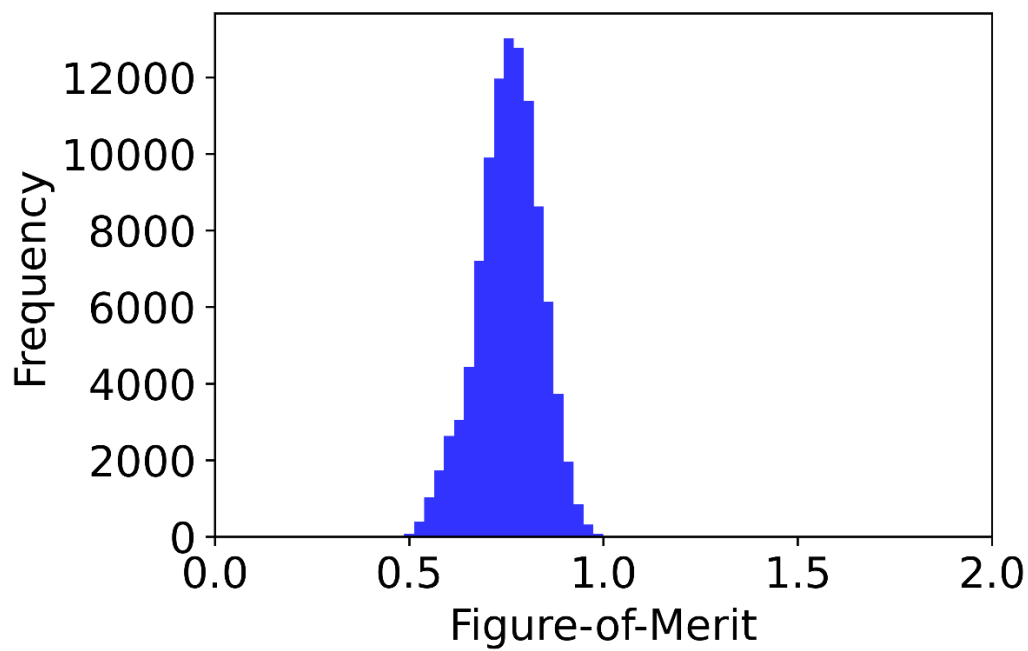

**Figure S12.** Distribution of the calculated figure-of-merit (larger values correspond to closer proximity with a Pareto-optimal RHEA).